\documentclass[aps,prb,preprint,groupedaddress,showpacs]{revtex4}
\usepackage[utf8]{inputenc}
\usepackage[english]{babel}
\usepackage{amsmath}
\usepackage{amsfonts}
\usepackage{amssymb}
\usepackage{graphicx}
\usepackage{hyperref}
\usepackage{hyperref}
\usepackage{wrapfig}

\usepackage[rightcaption]{sidecap}

\newcommand{\mf}{{m_{\rm f}}}
\newcommand{\mfsq}{{m^2_{\rm f}}}
\newcommand{\ma}{{m_{\rm a}}}
\newcommand{\masq}{{m^2_{\rm a}}}

\newcommand{\I}{{\rm i}}
\begin{document}

\title{Emerging mechanisms of magnetocaloric effect in phase-separated metals}

\author{V.V.~Ivchenko$^{1,2}$, P.A.~Igoshev$^{1,2}$}

\affiliation{$^1$Institute of Metal Physics, 620108, Kovalevskaya str. 18, Ekaterinburg, Russia\\
$^2$Ural Federal University, 620002, Mira str. 19, Ekaterinburg, Russia
}
\date{\today}
\begin{abstract}
We present a study of the magnetocaloric effect in metallic systems exhibiting first order magnetic transitions and focus on consequences of magnetic phase separation. We account for ferrimagnetic, ferromagnetic and Neel antiferromagnetic order. Based on the archetypal Hubbard model being treated within the mean-field approximation, we provide and explore its implications on the field-induced entropy change in metallic system with phase separation. Chosen framework allows us to properly analyze phase volumes' dependence on parameters of phase-separated (PS) system. Moreover, an account for phase separation boundaries as functions of magnetic field provides a natural splitting of the PS region, where each subregion corresponds to a different temperature dependence of entropy change: moving from one subregion to the other produces a kink, followed by a strong linear growth of entropy change.
We encounter a second order magnetic transition from paramagnetic to antiferromagnetic phase in PS region that occurs for particular parameter values. Despite the fact that both phases have zero total magnetization, the transition has a strong impact on entropy change. 
\end{abstract}

\maketitle

\section{Introduction}\label{sec:intro}

The magnetocaloric effect (MCE) is defined by a change in entropy $S$ (temperature $T$) of the material subjected to a magnetic field $h$ in isothermal (adiabatic) conditions. Despite the fact that MCE was discovered in 1881~\cite{Warburg}, theoretical aspects of phase transition thermodynamics, the determination of optimal properties for higher MCE 
as well as choosing the best candidate for electron model (localized or itinerant) are still under research that is of utmost fundamental and practical importance~\cite{DEOLIVEIRA201089}. 
Materials with a considerable value of MCE can be used as building blocks of new devices for effective heating or refrigeration.

Besides materials with ordinary ferromagnetic (FM) order, a great interest is drawn towards systems, where a ferromagnetic  order coexists with antiferromagnetic (AFM) one. 
A coexistence can be achieved in two ways, they either join forming ferrimagnetic order, or sample spatially separates in two distinct phases. There are many materials exhibiting a ferrimagnetic order: Ni$_{45}$Co$_5$Mn$_{37}$In$_{13}$\cite{ferri_compound_inverse}, MnRhAs\cite{ferri_anom}, Ni$_{1.68}$Co$_{0.32}$Mn$_{1.20}$Ga$_{0.8}$\cite{strange_ferri_anom} or phase separation (PS): MnFeP$_{0.8}$Ge$_{0.2}$\cite{CARON20093559} (PM---FM phase separation), La$_{0.27}$Nd$_{0.40}$Ca$_{0.33}$MnO$_3$\cite{PS_material_anom} (FM---charge-order phase separation), Mn$_{0.99}$Cu$_{0.01}$As, Gd$_{5}$Ge$_{2.3}$Si$_{1.7}$\cite{PS_Mce}(FM---AFM phase separation). 

In the last decade, a correct way of determining $\Delta S$ using experimental data in systems with first order magnetic transition (FOMT) and phase separation (PS) has inspired a lot of debate. 
Whereas the conventional Maxwell relation works well for compounds exhibiting second order magnetic transition, it fails for FOMTs.
One reason is that phase volumes depend on temperature and magnetic field~\cite{PS_material_anom} and neglecting this dependency leads to spurious result~\cite{GMCE}, which was discussed in Refs.~\onlinecite{ferri_compound_inverse,wrong_Maxwell}. The same arguments diminish the direct use of Clausius-Clapeyron equation for FOMT~\cite{CC_Equal_MR}. 
Another reason is that separate phases inside PS have different response to a magnetic field. That is why a development of a suitable theory of MCE in FOMT region is still a major field of research \cite{maxwell?,PECHARSKY20093541,XU20153149,2013:Oliveira,otherGMCEstudy,1storder,1325142}.

There is a lot of experimental data that demonstrates direct MCE, when $\Delta S < 0, \Delta T > 0$. However, there exist materials that exhibit inverse MCE~($\Delta S > 0, \Delta T < 0$): MnRhAs~\cite{ferri_anom}, Co and Mn Co-Doped Ni$_2$MnGa~\cite{strange_ferri_anom}, Er$_2$Fe$_{17}$~\cite{KHEDR2019436,PhysRevB.86.184411}, DyAl$_2$\cite{PhysRevB.61.447}, Pr$_{0.46}$Sr$_{0.54}$MnO$_3$\cite{anomal_fail1st}, which can be connected with existing antiferromagnetic order and FOMTs. There are typically two distinct temperature intervals exhibiting inverse and direct MCE. Besides, an inverse MCE region demostrates a satellitelike peak, e.g.~Eu$_{0.55}$Sr$_{0.45}$MnO$_3$\cite{PS_material_anom},  Ni$_{50}$Mn$_{33.13}$In$_{13.90}$\cite{inverse_direct_PRB},ErGa$_2$, HoGa$_2$\cite{inverse_direct_anisotropy},  HoFeSi\cite{successive_anom_direct_1st}.

A localized electron model is one of the basic models for theoretical description of MCE, see applications in~Ref.~\onlinecite{DEOLIVEIRA201089}, which can be used for systems with a well-defined integer or half-integer-spin local moments (e.g. insulators).Also, the localized model can be treated as an effective model for metallic magnetism, on the~theoretical basis for exchange integral derivation in Ref.~\onlinecite{LIECHTENSTEIN198765} (e.g.~an application determining effective exchange integrals for $\alpha$-Fe is developed in Ref.~\onlinecite{2015:Igoshev.a-Fe}). Another application of the localized electron model can be found in Ref.~\onlinecite{von_Ranke_2009}, where authors achieve different types of magnetic order (ferromagnetic, ferrimagnetic, antiferromagnetic) by varying the parameters of the Hamiltonian
Antiferromagnetic instability may serve as a reason for a 
large inverse MCE~\cite{XIAO2018916}. Another mechanism for inverse MCE was due to strong anisotropy induced via interaction with crystal electric field\cite{PhysRevB.58.14436, CEF_anom_exp}.
However, metallic systems in FOMT region always exhibit phase separation (``mixed states''), and localized electron model can not account for appearing inhomogeneities.

An effective Landau theory based on an itinerant electron model was applied to MCE in Co(S,Se)$_2$, Lu(Co,Al)$_2$, and Lu(Co,Ga)$_2$ compounds~\cite{2003:Yamada}. 
This approach was developed by modeling the density of $d$-states MnAs~\cite{2004:Oliveira,2013:Oliveira,DEOLIVEIRA201089}, MnFeP$_{0.45}$As$_{0.55}$~\cite{2005:Oliveira} and by accounting for magnetoelastic interaction that changes the electron band structure with the change of magnetization. 
FOMT to ferromagnetic state and metamagnetic transitions were explained within Landau theory in terms of itinerant electron model for YCo$_2$\cite{1975:Bloch,1992:Duc}, through the negative sign of the fourth order coefficient of the free energy expansion. In the above papers, the reason for FOMT to ferromagnetic state was due to negative sign of a second derivative of density of states (DOS) at the Fermi level: a change of parameters led paramagnetic state to hop discontinuously into ferromagnetic one.
However, one can consider a magnetic transition from ferromagnetic to antiferromagnetic state, which is always of first order without any further conditions.

From a practical point of view, one can have an interest not only in a large peak-like $\Delta S$, but also in other types of $\Delta S$ behavior that can find its technological applications. For example, the table-like temperature dependence of $\Delta S$ is one of such types and it can be used for constructing an ideal Ericsson cycle of magnetic refrigeration. Recently the compounds exhibiting such behavior attracted a lot of attention~\cite{2014:Ericsson:Gd-Co-Al,2015:Ericsson:Gd56Ni15Al27Zr2,2016:Ericsson:Gd50Co45Fe5,2017:Ericsson:Gd-Ni-Al,2018:Ericsson:GdMnSi}. 
While the origin of table-like behavior of $\Delta S$ is typically the amorphousness or heterogeneity of alloys or compounds under consideration, the origin can have intrinsic property of the phase transition provided that this is affected by phase-separation effects.  




This paper is organized in the following way. In Sec.~\ref{sec:schema} 
we present purely qualitative description of phase separation providing a schematic representation for the phase-separated region on phase diagram,
in Sec. \ref{sec:model} we describe the model and methods accounting for quantitative treatment of antiferromagnetism and non-homogeneous states. In Sec. \ref{sec:results} we discuss lattice choice and present calculated phase diagrams, including ferromagnetic, antiferromagnetic, as well as non-homogeneous and ferrimagnetic states. After that we present a few examples of calculating $\Delta S$ in the vicinity of the tricritical point and discuss the obtained results. In Sect.~\ref{sec:conclusions} we conclude the paper.

\section{Qualitative perspective on phase separation}\label{sec:schema}

The main distinction between itinerant electron and Heisenberg-like models is that in the former the electron transfer leads to phase separation at FOMT point: system splits in two distinct phases, which can have different response to a magnetic field, and energy equilibrium is achieved by equal phase-energies per particle. Such condition is completely different from the conventionally used approach based on Heisenberg-like model where phase-energies per site should be considered equal. Despite the fact that this leads to a rather different thermodynamic properties, it eludes attention in MCE studies.

\begin{figure}
    \centering
    \includegraphics[width=0.5\linewidth]{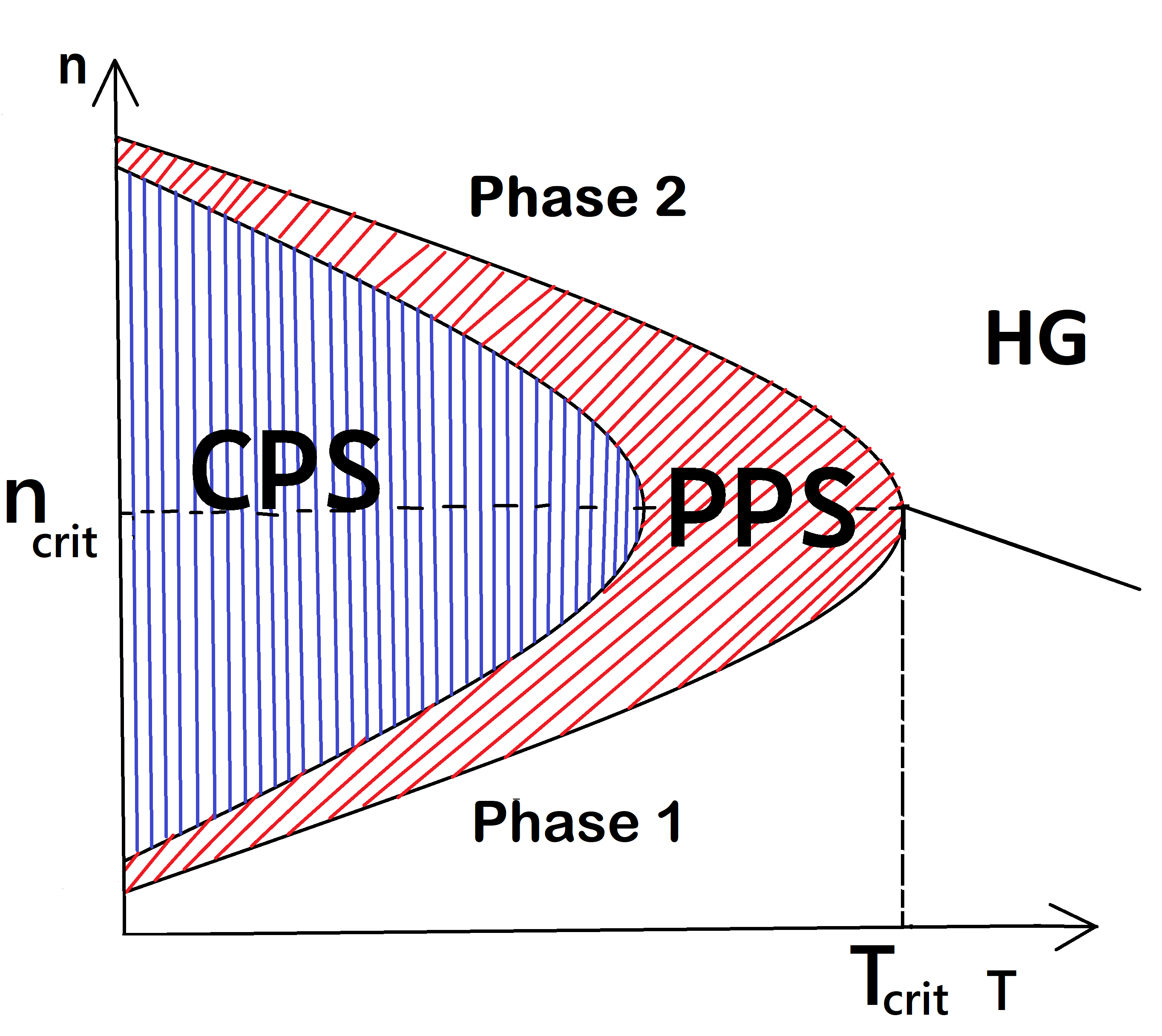}
    \caption{(Color online) Schematic representations of a vicinity of a tricritical point $(n_{\rm crit},T_{\rm crit})$ on the phase diagram depicted in terms of temperature($T$) --- filling($n$): CPS --- phase separation region ``Phase 1 + Phase 2'' in both zero and finite magnetic field, PPS --- phase separation only exists in finite or zero magnetic field, HG, homogeneous phase (Phase 1 or Phase 2) in both zero and finite magnetic field. Black line corresponds to a second order transition between Phase 1 and Phase 2.}
    \label{fig:general}
\end{figure}

Here we generalize and imply the approach of~Ref.~\onlinecite{Igoshev:2010,Igoshev_2015}, previously developed for the investigation of magnetic phase separation in the ground state of the Hubbard model for a consistent study of MCE for the electron system exhibiting FOMT. 
Qualitative analysis of phase separation properties leads one to consider schematic picture of phase diagram in terms of temperature ($T$) and electron filling ($n$). The most important region is the vicinity of tricritical point~$(n_{\rm crit},T_{\rm crit})$, Fig.~\ref{fig:general}, which is a point where phase transition changes its order from first to second (PS filling boundaries merge).
At a given value of magnetic field, the vicinity of tricritical point can naturally be split in three distinct subregions: CPS (complete phase separation) where PS exists in both zero and non-zero values of magnetic field; PPS (partial phase separation) where PS exists only in $h = 0$ or $h \neq 0$; HG (homogeneous) where there is no phase separation. For given splitting, we expect entropy change to exhibit non-trivial temperature and magnetic field dependence. PS boundaries adjust correspondingly to an application of magnetic field and PPS-subregion is formed. Since phase separation implies an existence of two distinct phases with their own magnetic response, one has to introduce phase volumes.
At fixed filling $n$, phase volume changes with temperature and approaching PPS (HG) achieves $0$ or $1$, depending on the relative position of $n$ with respect to~$n_{\rm crit}$. 
Increasing $T>T_{\rm crit}$ with $n < n_{\rm crit}$ ferromagnetic component orientation results in full vanishing of the PS-region and a domination of the second order phase transition~(SOMT). 
For a fixed $T$ increase in filling $n$ leads to continuous transition of Phase $1$ into PS, where system separates in two independent phases Phase 1 + Phase 2, which have different responses to a change in magnetic field. Note, that inside PS a change in $n$ is only reflected on phase volumes, while properties of the phases stay the same.

\section{Model and approximations}\label{sec:model}


Non-degenerate Hubbard model is chosen as a simplest model to account for the impact of phase separation in metallic (itinerant) electron systems. This model allows an easy computation of AFM ordering parameters.

The Hubbard model Hamiltonian with magnetic field $h$ directed along $z$-axis has the a form
\begin{equation}\label{eq:Hubbard_model}
	\mathcal{H} = \sum_{\mathbf{k}\sigma}(\epsilon_{\mathbf{k}} - \gamma_{\sigma}h )c_{\mathbf{k} \sigma}^\dagger c_{\mathbf{k}\sigma} + U \sum_{i}n_{i\uparrow }n_{i\downarrow},
\end{equation}
where $c^\dagger_{\mathbf{k} \sigma}/c_{\mathbf{k} \sigma}$ are electron creation/annihilation operators, $n_{i\sigma} = c^\dag_{i\sigma}c_{i\sigma}$ is a particle-number operator, $c_{\mathbf{k}\sigma} = N^{-1/2}\sum_i\exp(\I\mathbf{kR}_i)c_{i\sigma}$, $N$ is a number of sites, $\mathbf{R}_i$ is a vector-valued position of a site in a lattice, $\sigma$ is a spin projection on the $z$-axis, $\gamma_{\sigma} = +1(-1)$ where $\sigma = \uparrow(\downarrow)$.

We choose the mean-field (generalized Hartree-Fock) approximation for ferro-, antiferromagnetic and their imposition in the same and perpendicular directions: ferrimagnetic and canted antiferromagnetic order, respectively~\cite{PhysRev.142.350}.

Ferrimagnetic (``Fi'') case is phase, where components of magnetic order are collinear, 
\begin{equation}
\label{eq:m_Fi_def}
\mathbf{m}^{\rm Fi}_i = [\mf + \exp(i \mathbf{QR}_i] \ma)\mathbf{e}_z,
\end{equation}
and canted antiferromagnetic (``CA'') phase, in which ferromagnetic orientation is perpendicular to antiferromagnetic,
\begin{equation}
\label{eq:m_CA_def}
\mathbf{m}^{\rm CA}_i =  \mf\mathbf{e}_z + \exp(i \mathbf{QR}_i) \ma\mathbf{e}_x,
\end{equation}
where $\mathbf{e}_i$ are unit vectors in Cartesian coordinates, $\mathbf{Q}$ is antiferromagnetic wave vector (for bipartite lattice $\exp(\I\mathbf{QR}_i) = \pm1$ depending on what AFM sublattice $i$th site is located~in), $\mf$ and $\ma$ are ferromagnetic and antiferromagnetic (staggered) components of the magnetization, respectively. 
                                                                                                                     
In order to account for AFM phase we treat the interaction term in Eq.~(\ref{eq:Hubbard_model})
\begin{equation}\label{eq:decoupling}
U \sum_{i}n_{i\uparrow}n_{i\downarrow} = \frac{U}4 \sum_{i}(n_i^2 - \mathbf{m}_i^2) \rightarrow \frac{U}2 \sum_{i}(n_i n - \mathbf{m}_i \langle \mathbf{m}_i \rangle) 
- UN n^2/4 + \frac{U}4\sum_{i}\langle \mathbf{m}_i \rangle^2,
\end{equation}
where
$n_i = \sum_{\sigma}c_{i \sigma}^\dagger c_{i \sigma}$ and $\mathbf{m}_i = \sum_{\sigma \sigma^{'}}c_{i \sigma}^\dagger \vec{\sigma}_{\sigma \sigma'} c_{i\sigma'}$, where $\vec{\sigma}$ is Pauli matrix vector.
Statistical average values of the above operators are 
\begin{equation}
\langle n_i \rangle = n,
\label{eqn:1}
\end{equation}
and two possibilities: 
\begin{equation}\label{eq:magnetic_order}
\langle \mathbf{m}_i \rangle = 
\begin{cases}
\mathbf{m}^{\rm Fi}_i, \mbox{Fi-order(ing)},\\
\mathbf{m}^{\rm CA}_i, \mbox{CA-order(ing)},
\end{cases}
\end{equation}
where $\mathbf{m}^{\rm Fi}_i$, $\mathbf{m}^{\rm CA}_i$ are defined by~Eqs.~(\ref{eq:m_Fi_def}) and~(\ref{eq:m_CA_def}). 

Substituting Eqs. (\ref{eq:m_Fi_def}),(\ref{eq:m_CA_def}) in Eq. (\ref{eq:decoupling}) we get the effective Hamiltonian in the mean-field approximation. A diagonalization of this yields two branches of the spectrum ($s = 1,2$)
\begin{equation}\label{eqn8}
 E_{s\sigma}^{\mathrm{Fi}}(\mathbf{k}) = e^+_{\mathbf{k}} - \gamma_{\sigma} \Delta_{\rm f} + Un/2 + (-1)^{s}\sqrt{\left(e^-_{\mathbf{k}}\right)^2 + \Delta_{\rm a}^2},  
\end{equation}
for ferromagnetic order (``Fi''), see~Eq.~(\ref{eq:m_Fi_def}), and spectrum
\begin{equation}
    E_{s\sigma}^{\mathrm{CA}}(\mathbf{k}) = e^+_{\mathbf{k}}+ Un/2 + (-1)^s\sqrt{(e^-_{\mathbf{k}\sigma})^2 + \Delta^2_{\rm a}},
    \label{spectre}
\end{equation}
in the case of incommensurate antiferromagnetic order (``CA''), see Eq.~(\ref{eq:m_CA_def}). Here $e_{\mathbf{k}\sigma}^{\pm} = e_\mathbf{k}^{\pm} - \gamma_{\sigma}\Delta_{\rm f}, e_{\mathbf{k}}^{\pm} = \dfrac{\epsilon_{\mathbf{k}}\pm \epsilon_{\mathbf{k+Q}}}{2}$, $\Delta_{\rm f} = U \mf/2 + h, \Delta_{\rm a} = U\ma/2$.

From here on in we consider the case of nesting of electronic spectrum, which means that $\epsilon_{\mathbf{k}} = - \epsilon_{\mathbf{k+Q}}$, since it works for bipartite lattice in the nearest-neighbor approximations~(it is enough to demonstrate influence of AFM order and phase separation on MCE). 
Direct calculation of statistical average $\langle\mathbf{m}_i\rangle$ in Eq.~(\ref{eq:magnetic_order}) yields the respective equations (for ``Fi'')
\begin{eqnarray}
  \label{eqn10}
    n &=& \dfrac{1}{2N}\sum_{s \sigma}\sum_{\mathbf{k}}f[E_{s\sigma}^{\mathrm{Fi}}(\mathbf{k})],\\
    \label{eqn11}
    \mf &=& \dfrac{1}{2N}\sum_{s \sigma}\sum_{\mathbf{k}} \gamma_{\sigma}  f[E_{s\sigma}^{\mathrm{Fi}}(\mathbf{k})],\\
    \label{eqn12}
    \ma &=& -\dfrac{\Delta_{\rm a}}{2N}\sum_{s \sigma}\sum_{\mathbf{k}} \frac{(-1)^s   f[E_{s\sigma}^{\mathrm{Fi}}(\mathbf{k})]}{\sqrt{\epsilon_{\mathbf{k}}^2 + \Delta_{\rm a}^2}},
\end{eqnarray}
where $f[E] = 1/(\exp[(E-\mu)/T] + 1)$ is the Fermi function.
And analogously in ``CA'' case we obtain
\begin{eqnarray}
    \label{ca10}
    n &=& \dfrac{1}{2N}\sum_{s \sigma}\sum_{\mathbf{k}}f[E_{s\sigma}^{\mathrm{CA}}(\mathbf{k})],\\
    \label{ca11}
    \mf &=&  \frac1{2N}\sum_{s \sigma}\sum_{\mathbf{k}} \dfrac{(-1)^{s}f[E_{s \sigma}^{\mathrm{CA}}(\mathbf{k})]\left(\epsilon_{\mathbf{k}} - \gamma_{\sigma}\Delta_{\rm f} \right)}{\sqrt{\left(\epsilon_{\mathbf{k}} - \gamma_{\sigma}\Delta_{\rm f} \right)^2 + \Delta_{\rm a}^2}},\\
    \label{ca12}
    \ma &=& -\frac{\Delta_{\rm a}}{2N} \sum_{s \sigma}\sum_{\mathbf{k}}  \dfrac{(-1)^{s}f[E_{s \sigma}^{\mathrm{CA}}(\mathbf{k})]}{\sqrt{\left(\epsilon_{\mathbf{k}} - \gamma_{\sigma}\Delta_{\rm f} \right)^2 + \Delta_{\rm a}^2}}.
\end{eqnarray}
These equations reproduce as a particular case the equation on Curie temperature $T_{\rm C}$ of the Stoner theory (in the limit case $\mf\rightarrow 0$, $\ma = 0$) 
\begin{equation}
\dfrac{U}{N}\sum_{\mathbf{k}}f'[\epsilon_{\mathbf{k}} + Un/2] + 1 = 0
\end{equation}
and on N\'eel temperature $T_{\rm N}$ of the Overhauser theory (in the limit case $\ma\rightarrow 0$, $\mf = 0$) 
\begin{equation}
\dfrac{U}{N}\sum_{\mathbf{k}}f[\epsilon_{\mathbf{k}} + Un/2]/\epsilon_{\mathbf{k}} + 1 = 0.
\end{equation}
\begin{figure}[b]
    {\includegraphics[width=.48\linewidth]{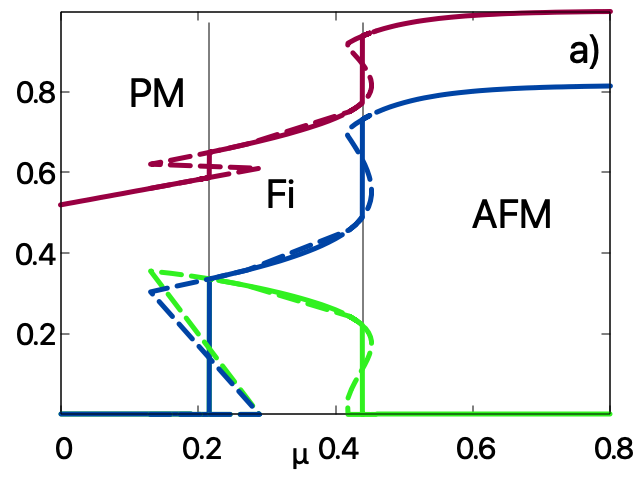}}
    {\includegraphics[width=.48\linewidth]{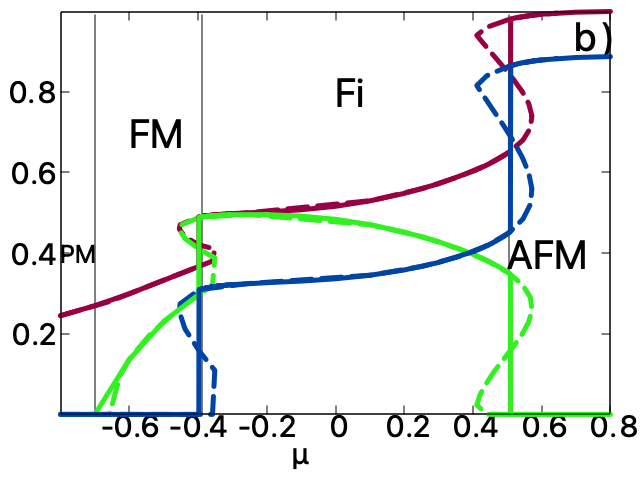}}
    \caption{(Color online) Example of $n$ (red line), $\mf$~(green line), $\ma$~(blue line) dependence on chemical potential $\mu$ in zero magnetic field and $T = 0.1$. (a) $U = 3$, (b) $U = 4$. Solid (dashed) lines: with (without) an account for phase separation in~FOMT. For the dashed lines we see violation of thermodynamic relation $\partial n/\partial\mu > 0$. For filled lines we see jumps of all functions. The jump for $n$ gives the difference in electronic filling for the phases taking part in the phase separation.}
    \label{fig:PS_demonstration}
\end{figure} 
 
Thus, a grand potential within the Hartree-Fock approximation reads
\begin{equation}\label{eq:Omega_def}
    \Omega = \frac{NU}4(\mfsq + \masq) - \dfrac{T}{2} \sum_{\mathbf{k}s\sigma}\ln\left(1 + \exp\left(\dfrac{E_{s\sigma}(\mathbf{k}) - \mu}{T}\right) \right),
\end{equation}
where $E_{s\sigma}(\mathbf{k})$ is the chosen electron spectrum, $\mu$ is a chemical potential. 

Here we go from the sum in $\mathbf{k}$ to an integral using DOS of the bare electronic states
\begin{equation}\label{eq:DOS_def}
    \rho(e) = \frac1{N}\sum_{\mathbf{k}}\delta(e - \epsilon_{\mathbf{k}})
\end{equation}
for the grand potential
\begin{equation}
\label{eqn:omega}
    \Omega = \dfrac{T}{2}\sum_{s \sigma} \int de \rho (e) K_{\Omega} \left(\dfrac{E_{s\sigma}(\mathbf{k}) - \mu}{T}\right) + \frac{U}4\left(\masq + \mfsq - n^2\right),
\end{equation}
where
\begin{equation}
    K_{\Omega}(z) = \ln[1 + \exp(-z)]
\end{equation}
and for entropy
\begin{equation}
\label{eqn:entropy}
    S = \dfrac{1}{2}\sum_{s \sigma} \int de \rho (e) K_{S} \left(\dfrac{E_{s\sigma}(\mathbf{k}) - \mu}{T}\right)
\end{equation}
where
\begin{equation}
    K_{S}(z) = \ln[1 + \exp(-z)] + \dfrac{z}{2}\tanh\dfrac{z}{2},
\end{equation}
In these equations $E_{s \sigma}(\mathbf{k}) = E_{s \sigma}^{\Phi}(\mathbf{k})$, where $\Phi = \mathrm{Fi}, \mathrm{CA}$. 
The solution of stationary point equations (\ref{eqn10})--(\ref{eqn12}), (\ref{ca10})--(\ref{ca12})  and Eq.~(\ref{eqn:omega}) allows to calculate the free energy $F = \Omega + \mu  n$ for each phase. The choice of the phase with minimal free energy indicates the most stable uniform state of the compound. 
Free energy of non-homogeneous states, which are formed in FOMT, is directly compared to energies of conventional phases (FM,PM,AFM) and phases with imposed ferro- and antiferromagnetic orders (ferrimagnetic or canted). It allows proper description of thermodynamics of phase transitions for parameters not only inside phase separation but on the outside as well and to calculate field-induced entropy changes.

Here we present a detailed description of the method that allows for convenient analysis of phase separation inside FOMT\cite{Igoshev:2010}. The main idea is to go from typically used pair of parameters $(n, F)$ to $(\mu, \Omega)$. Assume that for a particular $\mu$ there exist two solutions $n_1(\mu, T, h), \mathcal{M}_1(\mu, T, h)$ and $n_2(\mu, T, h), \mathcal{M}_2(\mu, T, h)$ of~Eqs.~(\ref{eqn10}),(\ref{eqn11}) and (\ref{eqn12}) or ~Eqs.~(\ref{ca10}),(\ref{ca11}) and (\ref{ca12}), corresponding to Phase 1 and Phase 2, respectively, which locally minimize $\Omega$, see Eq. (\ref{eqn:omega}). Here $\mathcal{M} = (m_{\rm f}, m_{\rm a})$ is a formal symbol for multicomponent order parameter. 
Let $\Omega(\mu, T, h | n_1, \mathcal{M}_1) < \Omega(\mu, T, h | n_2, \mathcal{M}_2)$ with $\mu < \mu_{\rm c}(T, h)$ and $\Omega(\mu, T | n_1, \mathcal{M}_1) > \Omega(\mu, T, h | n_2, \mathcal{M}_2)$ with $\mu > \mu_{\rm c}(T, h)$. 
With an increase in $\mu$, $n$ and $\mathcal{M}$ jumps when $\mu = \mu_{\rm c}$, and the states with $n_1 < n < n_2$ are not enacted as homogeneous. Within given approach, one can also achieve a description of SOMT by considering $n_1 = n_2$, and $\mathcal{M}_1 = \mathcal{M}_2$.

To describe a phase separation case, for a given $n$, we introduce phase volumes $x_1, x_2$ that depend on the parameters of the system via following relation
\begin{equation}
	\label{eq:x2}
	n = x_1n_1(\mu_{\rm c}(T, h),T,h) + x_2n_2(\mu_{\rm c}(T, h),T,h),	
\end{equation}
and $x_1 + x_2 = 1$, so phase volumes are linear functions of $n$: $x_1(n; n_1, n_2) = (n_2 - n)/(n_2 - n_1)$, $x_2(n; n_1, n_2) = (n - n_1)/(n_2 - n_1)$. Therefore, for any extensive quantity $A$ per sample site in phase-separated region we have ($a_1$ and $a_2$ are specific values of $A$, i.e. values per phase site), we have 
\begin{equation}
	\label{eq:m_x}
	A = x_1a_1(\mu_{\rm c}(T, h),T,h) + x_2a_2(\mu_{\rm c}(T, h),T,h),
\end{equation}
where $x_i a_i(\mu_{\rm c}(T, h),T,h) = A_i$ being phase value per sample site.
In particular for entropy $S$ we obtain
\begin{equation}
	\label{eq:S_x}
	S(n,T,h) = S_1(n,T,h) + S_2(n,T,h),
\end{equation}
where $S_i(n,T,h) = x_i(n; n_1(\mu_{\rm c}(T, h),T,h), n_2(\mu_{\rm c}(T, h),T,h))s_i(\mu_{\rm c}(T,h),T,h)$ where $S_i$ corresponds to phase entropy per sample site, and $s_i$ is a specific entropy.
In phase-separated region electron filling $n$ is completely determined by the phase volumes, whereas the phase properties remain to be unchanged.

Note, that phase volumes dependence on temperature and magnetic field is one of the reasons why the Maxwell relation 
\begin{equation}
\label{eq:Maxwell}
   \left(\dfrac{\partial S}{\partial h}\right)_{T} = \left(\dfrac{\partial M}{\partial T}\right)_{h},
\end{equation}
$M$ being a sample magnetization, fails inside PS, see~Ref.~\onlinecite{PS_material_anom}.

To demonstrate an account for phase separation, we provide examples of filling $n$ and magnetization amplitudes $\mf, \ma$ dependence on $\mu$ (for parameters used in a full calculation of MCE below), see Fig.~\ref{fig:PS_demonstration}. An algorithm for calculations is the following: for a fixed temperature and chemical potential the algorithm minimizes grand potential (\ref{eqn:omega}) with~respect to order parameters, which determine the phase of the system; if the next step in chemical potential has as its minimum a distinct phase from the previous step, then the algorithm launches a procedure of localization $\mu = \mu_{\rm c}$ corresponding to the phase change and thereby PS.

Metastable states, whose homogeneity is not stable, are not realized and are replaced by a mix of two phases, corresponding to parameters on the boundaries of the jump interval $(n_1, \mathcal{M}_1)$ and $(n_2, \mathcal{M}_2)$.

The above given procedure is completely general, it does not depend on the approximation and can be used for every lattice, which can be described via DOS $\rho(\epsilon)$. 

\section{Results}\label{sec:results}

\begin{figure}[t]
\includegraphics[width=0.48\linewidth]{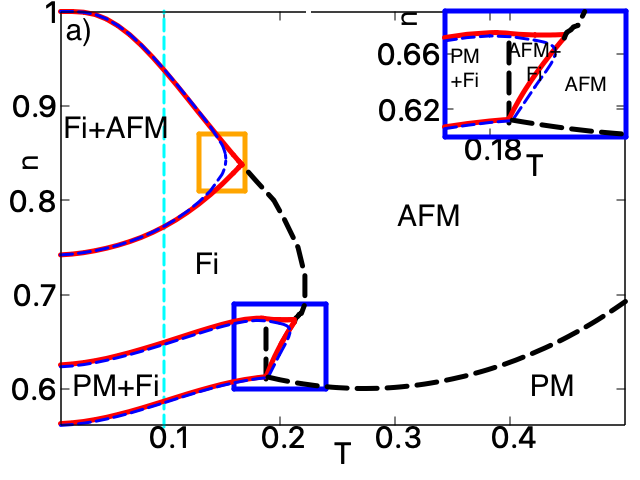}
\includegraphics[width=0.48\linewidth]{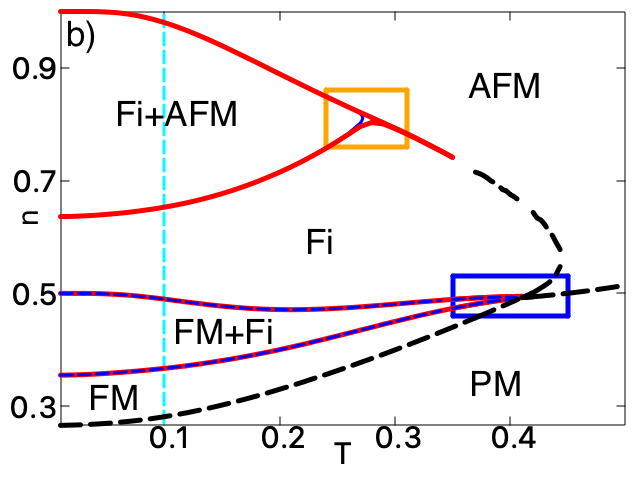}
    \caption{(Color online) Phase diagram in $T-n$ variables.  
    AFM is antiferromagnetic phase, Fi is ferrimagnetic phase ($\mf > 0, \ma > 0$), PM is paramagnetic phase, FM is ferromagnetic phase, $\Phi_1 + \Phi_2$ is PS region of $\Phi_1$ and $\Phi_2$. Red lines are PS boundaries in zero magnetic field, dashed black lines are phase boundaries of second order magnetic transition. Blue dashed lines are phase separation boundaries in magnetic field $h$. Parameters (a) $U = 3$, $h  = 10^{-3}$, (b) $U = 4$, $h  = 10^{-4}$.
    Cyan lines correspond to a temperature, for which the calculations were done in Fig.~\ref{fig:PS_demonstration}. Rectangle regions are vicinities of tricritical points for which the calculations below were done.}
    \label{fig:pic1}
\end{figure}

To illustrate a FOMT influence on MCE in an electron subsystem of solid we choose a density of states of the Bethe lattice with nearest-neighbor hopping approximation in the limit of infinite neighbors\cite{Bethe} 
\begin{equation}
    \rho(e) = \dfrac{1}{2 \pi} \sqrt{4D^2-e^2}.
\end{equation}
Here and below we take $D$ as an energy unit. 
This choice of DOS approximately reproduces the behavior of DOS for simple cubic lattice within the nearest neighbor-hopping approximation, excluding van Hove singularities.

We use equations from Sec. \ref{sec:model} to compare grand potential of ferromagnetic, antiferromagnetic, ferrimagnetic and canted antiferromagnetic phases and to choose the minimal one (see Ref.~\onlinecite{Igoshev:2010}) which allows to calculate phase diagrams in $n-T$ plane including non-homogeneous states (phase separation).


The following parameters were chosen: (i) $U = 3, h = 10^{-3}$ , (ii) $U = 4, h = 10^{-4}$, which correspond to the values of Coulomb interaction and magnetic field so FOMTs between ferromagnetic (ferrimagnetic) and antiferromagnetic phases become possible. The value of magnetic field $h$ for $U = 4$ was chosen to be lower in order to compensate for a rapid growth of the PPS-subregion with increase of $U$. At first, an increase in Coulomb interaction introduces ferrimagnetic order ($U = 3$), and then ferromagnetic order ($U = 4$). For all considered parameter values, ``CA'' phase appears to be energetically unfavourable.

For $U = 3$ [Fig.~\ref{fig:pic1}a] we see that in sufficiently low temperatures $T < 0.22$ first order transition between paramagnetic and ferrimagnetic phases, and between ferrimagnetic and antiferromagnetic phases with increase of $n$ is accompanied by phase separation. Low values of $n$ correspond to paramagnetic phase. 


Phase diagram appears to have two tricritical points: for Fi--AFM phase transition [$T_{\rm crit} = 0.17, n_{\rm crit} = 0.84$, orange frame in Fig.~\ref{fig:pic1}(a)]; the analogous point for AFM--Fi phase transition~[$T_{\rm crit} = 0.22, n_{\rm crit} = 0.67$, blue frame in Fig.~\ref{fig:pic1}(a)].

Phase diagram for $U = 4$ is shown in Fig.~\ref{fig:pic1}(b). We observe not only an obvious expansion of magnetically-ordered region, but that PS region ``PM + Fi'' is now replaced by PS region ``FM + Fi'' and a new second order transition from PM to FM phase occurs. Phase diagram appears to have two tricritical points: for Fi--AFM phase transition~[$T_{\rm crit} = 0.28, n_{\rm crit} = 0.81$, orange frame in Fig.~\ref{fig:pic1}(b)]; for FM--Fi phase transition [$T_{\rm crit} = 0.41, n_{\rm crit} =0.49$, blue frame in Fig.~\ref{fig:pic1}(b)].


General perspective points out that magnetic phase separation should lead to noticeable changes in MCE. Different values of magnetic field correspond to different PS boundaries, which lead to a formation of PPS-subregion (Fig.~\ref{fig:general}). Therefore, instead of one critical temperature, we analyze two, namely the temperature of exit from CPS $T_{\rm CPS}(h,n)$ and PPS $T_{\rm PPS}(h,n)$, respectively. 

Inside PS, a change in $n$ changes phase volumes $x_1, x_2$ of respective phases, see Eq.~(\ref{eq:x2}). In the vicinity of tricritical point they represent major regulators of $\Delta S$, because below $n_{\rm crit}$ (Fig.~\ref{fig:general}), while approaching $T_{\rm PPS}$ ($T_{\rm {CPS}}$) main contribution to $\Delta S$ is given by the bottom phase, above $n_{\rm crit}$ --- by the top one.


\begin{figure}[t]
\centering
\includegraphics[width=0.5\linewidth]{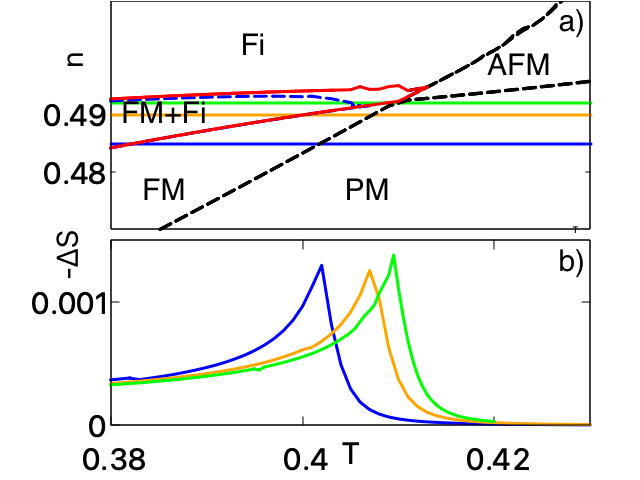}
\caption{(Color online) Region from blue frame from Fig.~\ref{fig:pic1}b; a: horizontal lines: $n = 0.48, n = 0.49, n = 0.492$; b: $\Delta S(T)$, colors correspond respectively.}
\label{fig:dS_U=4_lower}
\end{figure}

In the following part we investigate $\Delta S$ for parameters in the vicinity of the tricritical points. In these calculations, we follow the procedure given in Sec. \ref{sec:model}. In each case we analyze a few filling values in the vicinity of $n_{\rm crit}$.

Results for calculated temperature dependence of $\Delta S$ for $U = 4$ in PS region ``FM+Fi'' [blue frame, Fig.~\ref{fig:pic1}(b)] are shown in Fig.~\ref{fig:dS_U=4_lower}. We observe MCE peak that corresponds to a second order magnetic transition (SOMT). Since both phases in PS have non-zero ferromagnetic component, changes in magnetic field do not lead to any noticeable effect. In other words, formation of PPS-subregion is reduced and no substantial PS effect on $\Delta S$ is encountered.

\begin{figure}[b]
    \includegraphics[width=0.42\linewidth]{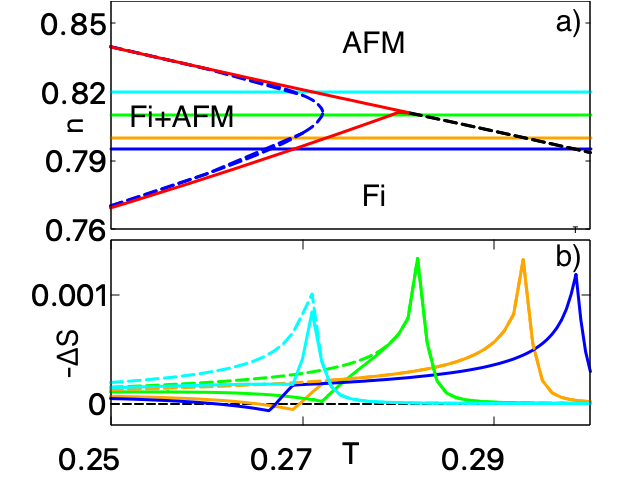}
    \includegraphics[width=0.47\linewidth]{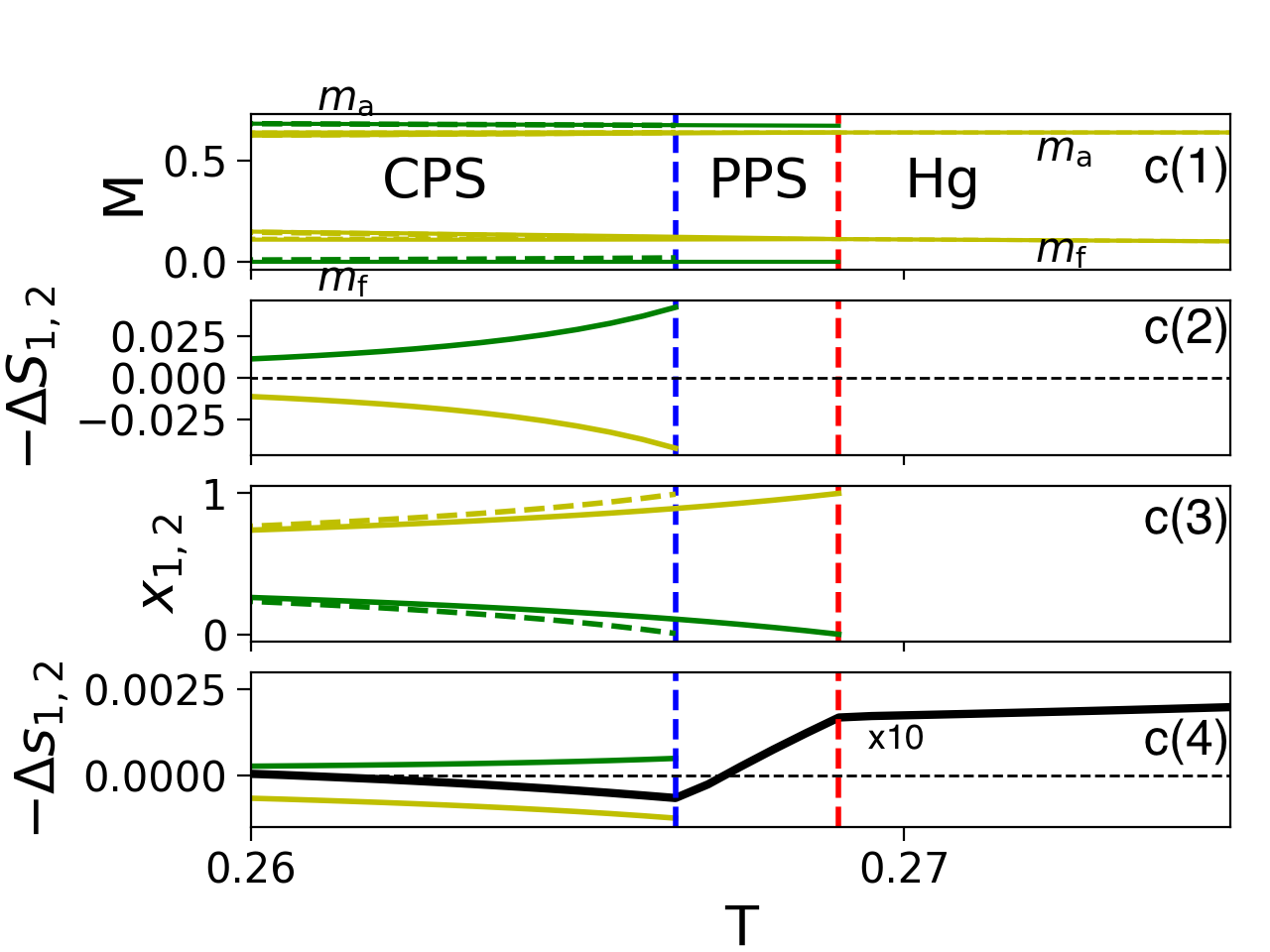}
    \caption{(Color online) $U = 4$. (a) Phase diagram corresponding to orange frame in Fig.~\ref{fig:pic1}(b); horizontal lines: $n = 0.795, n = 0.80, n = 0.81, n = 0.82$. (b) $\Delta S(T)$ for a fixed filling, line colors correspond to horizontal lines above. Dashed lines represent calculations without accounting for phase separation. (c) Temperature dependence of phase quantities for $n = 0.795$. Green lines correspond to AFM phase, yellow lines correspond to ferrimagnetic phase. Blue vertical dashed line corresponds to transition in homogeneous state in finite magnetic field, red vertical dashed line corresponds to transition in homogeneous phase in $h = 0$. (c1) Magnetization amplitudes, dashed lines correspond to finite magnetic field; (c2) $\Delta S_{\rm AFM,Fi}$, Eq.~(\ref{eq:S_x}); (c3) phase volumes $x_{\rm AFM}, x_{\rm Fi}$, dashed lines correspond to finite magnetic field; (c4) black color $\Delta S \times 10$, and differences in specific entropies $\Delta s_{\rm AFM,Fi}$.}
    \label{fig:dS_U=4_upper}
\end{figure}

We proceed to analysis of the temperature dependence of $\Delta S$ and the corresponding phase contributions for $U = 4$, where system exhibits separation ``Fi+AFM'' [Fig.~\ref{fig:pic1}(b), orange frame], shown in Fig.~\ref{fig:dS_U=4_upper}. We present results with and without (dashed lines) an account for phase separation. It is evident that phase separation drastically alters the behavior of $\Delta S$. We discover one-sided peak with $\Delta S > 0$ exactly at the point $T_{\rm{CPS}}(h,n)$. An increase in filling lowers absolute value of the one-sided peak. 
Since filling $n = 0.795$ has large inverse MCE, we focus our investigation on that particular value of $n$, analyzing phase contributions in $\Delta S$, magnetizations and phase volumes, see Fig.~\ref{fig:dS_U=4_upper}(c). From Fig.~\ref{fig:dS_U=4_upper}(c2) we can see that $\Delta S_{\rm{AFM}}$ and $\Delta S_{\rm{Fi}}$ are strongly dependent on temperature. Therefore, in CPS-region we see strong reaction of the individual phases in PS to a change in magnetic field. Such effect appears to be due to phase volumes dependence on temperature, see Fig.~\ref{fig:dS_U=4_upper}(c3). In particular, the phase-balance inside PS increases typically small response of AFM phase to a magnetic field. Since $\Delta S_{\rm{Fi}}$ has opposite sign but the same absolute value as  $\Delta S_{\rm{AFM}}$, they cancel each other resulting in almost no response of resulting entropy $S$ to magnetic field.

Temperature dependence of the specific entropy changes is shown in Fig.~\ref{fig:dS_U=4_upper}(c4). Whereas AFM phase demonstrates direct MCE, $\Delta s_{\rm{AFM}}< 0$, ferrimagnetic phase exhibits an opposite tendency, $\Delta s_{\rm{Fi}} > 0$, moreover $|\Delta s_{\rm{Fi}}|>|\Delta s_{\rm{AFM}}|$. Since $n = 0.795$ is below $n_{\rm crit}$ (Fig.~\ref{fig:general}), $x_{\rm{Fi}} > x_{\rm{AFM}}$ [Fig.~\ref{fig:dS_U=4_upper}(c3)], we see a reverse growth of $\Delta S$ ending in the one-sided peak. The greater phase volume of ferrimagnetic phase, the greater value of inverse MCE is encountered. And, conversely, an increase in filling leads to increase in AFM phase volume, which in its turn suppresses inverse contribution of Fi phase. 
Inverse MCE of ferrimagnetic phase is due to decreasing $\mf$ on application of magnetic field, Fig.~\ref{fig:dS_U=4_upper}(c1), which is a rather general behavior: for a homogeneous system the relation $dm/dh < 0$ is strictly prohibited by fundamental equilibrium conditions~\cite{landau2013statistical}; however, for the phases inside PS it becomes possible. 

PPS-region exhibits a linear growth of $\Delta S$. However, with $n > 0.81$, the growth ends at the point $T_{\rm{PPS}}(h,n)$, while for fillings $n \leq 0.81$ in HG-region we observe SOMT from ferrimagnetic to AFM phase to which MCE peak is associated. Figure~\ref{fig:dS_U=4_upper}(c4) demonstrates a correlation between temperature interval of PPS-subregion and the region of linear growth of $\Delta S$. Therefore, if PPS-subregion is large enough and phases in PS substantially differ in their magnetic properties, we see drastic changes in $\Delta S$ behavior (in Fig.~\ref{fig:dS_U=4_upper} in PPS-region separation exists only in zero magnetic field).

\begin{figure}[t]
    \includegraphics[width=0.42\linewidth]{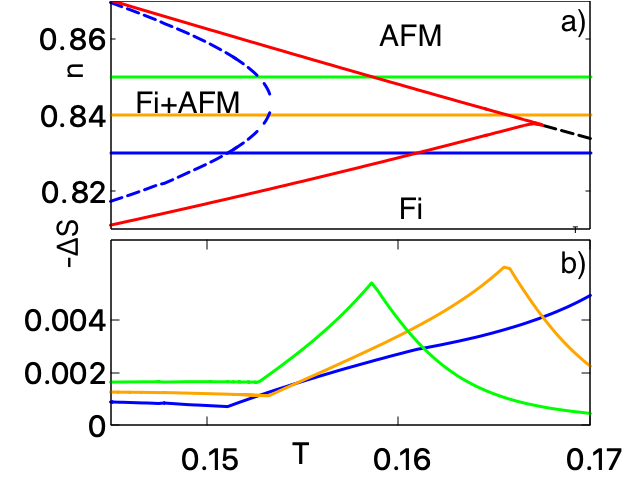}
    \includegraphics[width=0.47\linewidth]{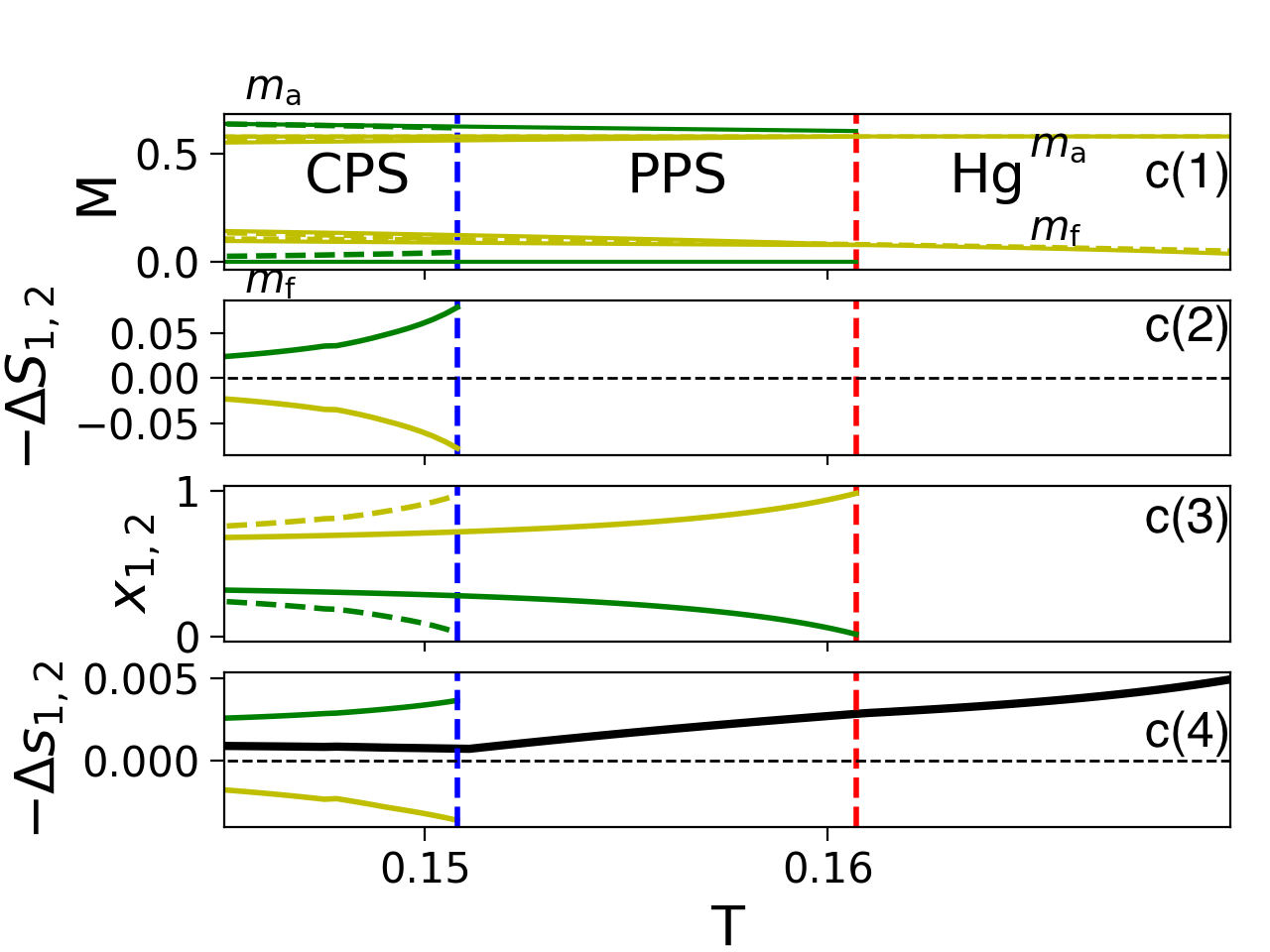}
    \caption{(Color online) $U = 3$. (a) Phase diagram corresponding to orange frame in Fig.~\ref{fig:pic1}a; horizontal lines: $n = 0.83, n = 0.84, n = 0.85$. (b) $\Delta S(T)$ for a fixed filling, colors correspond to horizontal lines in Fig.~\ref{fig:dS_U=3_upper}a. (c) Phase quantities for $n = 0.83$. Color and line type conventions are inherited from above (Fig.~\ref{fig:dS_U=4_upper}).}
    \label{fig:dS_U=3_upper}
\end{figure}

Results for $\Delta S(T,n)$ for $U = 3$ in the vicinity of tricritical point, where system exhibits separation ``Fi+AFM'' [orange frame in Fig.~\ref{fig:pic1}(a)], are shown in Fig.~\ref{fig:dS_U=3_upper}.
In CPS-subregion we observe a weak suppression of $\Delta S(T, n)$; PPS-subregion corresponds to a linear growth of $\Delta S$; in HG-subregion we encounter two different scenarios, where the first is a strong decrease of $\Delta S$, and the other is a region where $\Delta S$ continues growing until it reaches the point of SOMT. Comparing results to an analogous region in Fig.~\ref{fig:dS_U=4_upper}(b), we observe a strong suppression of the one-sided peak, which correlates with a decrease in Coulomb interaction to $U = 3$ leading to a lesser impact of ferrimagnetic phase (this effect does not depend on the field value). 
The inverse MCE of ferrimagnetic phase [Fig.~\ref{fig:dS_U=3_upper}(c4)] is again due to decreasing $\mf$ upon application of magnetic field. Here, $\Delta s_{\rm{Fi}}$ has approximately the same absolute values as AFM phase, Fig.~\ref{fig:dS_U=3_upper}(c4), so no inverse MCE is encountered.
We focus our attention on results for $n = 0.83$, because this filling corresponds to large ferrimagnetic phase volume, which leads to a slight decrease in overall $\Delta S$ in CPS-subregion.

On~Fig.~\ref{fig:dS_U=3_upper}(c2) we again encounter equality of absolute values $\Delta S_{\rm{Fi}}$ and $\Delta S_{\rm{AFM}}$ but opposite in sign, which leads to small response of $S$ to an application of magnetic field. PPS-subregion is formed with $h = 0$, i.e.,~PS exists only in zero magnetic field.

\begin{figure}[t]
    \includegraphics[width=0.42\linewidth]{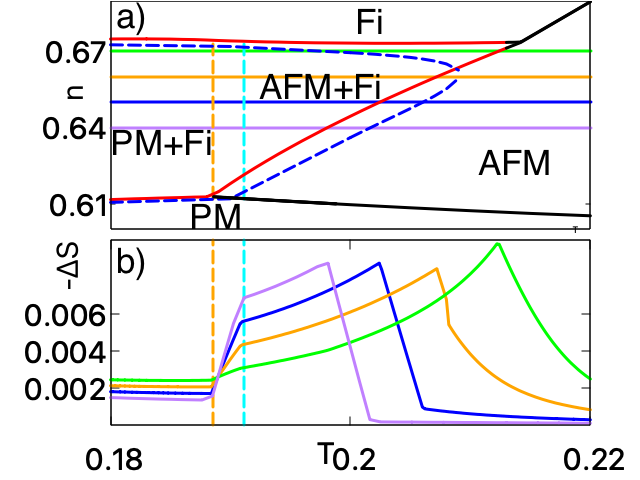}
    \includegraphics[width=0.47\linewidth]{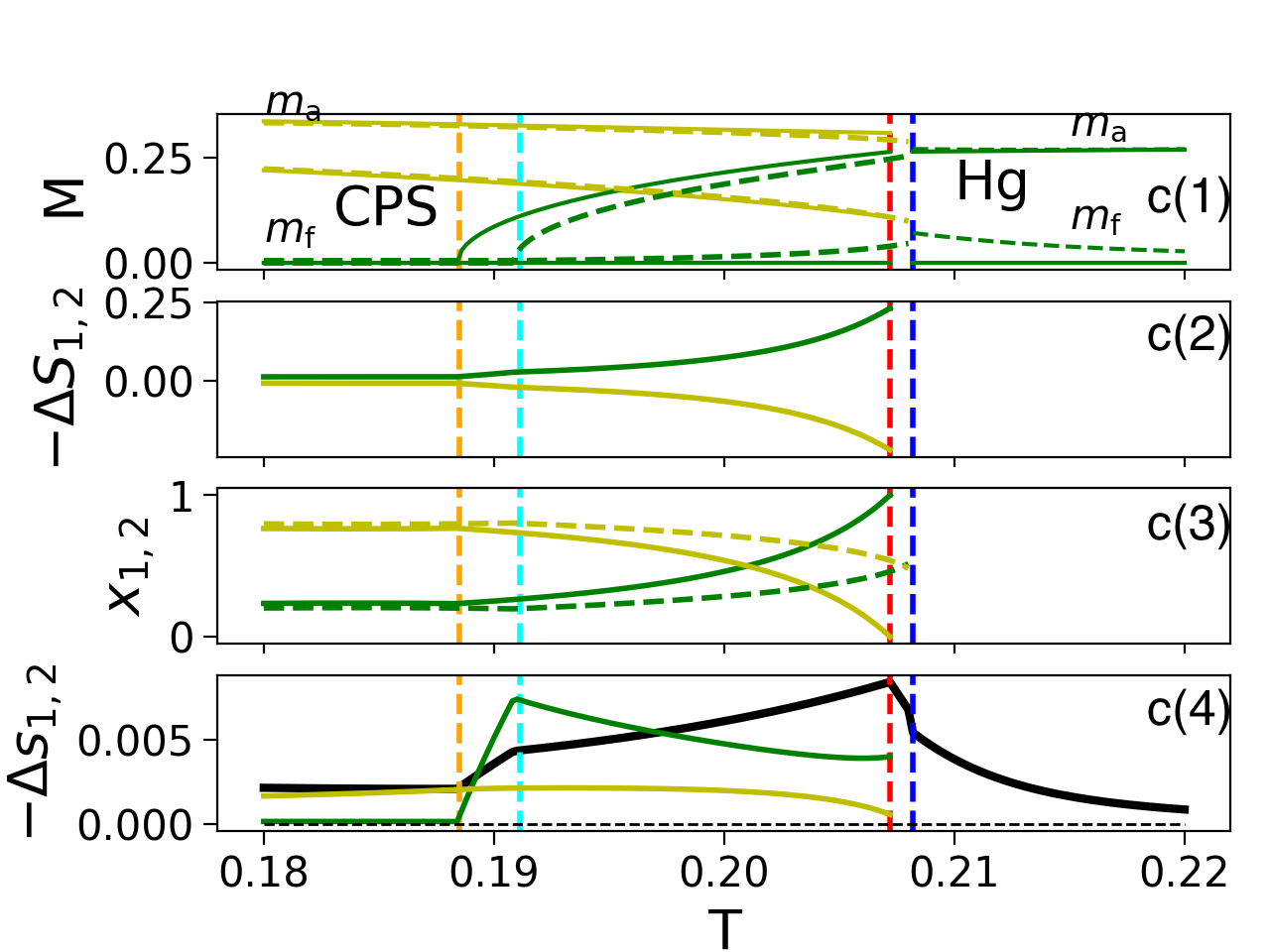}
    \caption{(Color online) $U = 3$. (a) Phase diagram corresponding to blue frame in Fig.~\ref{fig:pic1}a; horizontal lines: $n = 0.64, n = 0.65, n = 0.66, n = 0.67$. (b) $\Delta S(T)$ for a fixed filling, colors correspond to horizontal lines in Fig.~\ref{fig:dS_U=3_lower}(a). (c) Phase quantities for $n = 0.66$. Orange (cyan) vertical dashed lines correspond to PM $\rightarrow$ AFM transition in $h = 0$ ($h \neq 0$). Color and line type conventions are inherited from above (Fig.~\ref{fig:dS_U=4_upper}).}
    \label{fig:dS_U=3_lower}
\end{figure}

On Fig.~\ref{fig:dS_U=3_lower}, we present results for $U = 3$ in the vicinity of tricritical point $0.16 < T < 0.22$ from blue inset in Fig.~\ref{fig:pic1}(a). The phase diagram becomes more complex. With increasing temperature system exhibits SOMT inside phase-separated region from PM to AFM phase. Application of magnetic field shifts critical temperature of the transition (cyan vertical dashed line) relative to the zero field transition (orange vertical dashed line). 
A region between cyan and orange dashed lines corresponds to a rapid growth of $\Delta S$, because turning on a magnetic field induces transition inside PS of AFM phase to PM phase, which one can easily verify from magnetization curves in Fig.~\ref{fig:dS_U=3_lower}(c1). In this region we see how phase volumes regulate $\Delta S$. For $n = 0.64$ (purple line) phase volume of PM/AFM phase is larger than for $n = 0.67$ (green line), which is reflected on $\Delta S$: the greater phase volume of PM/AFM phase, the greater slope of $\Delta S$ we encounter.
An interesting note is that the lower the filling, the more table-like behavior of $\Delta S$ is displayed. It is directly seen that the second order magnetic transition inside PS leads to a big growth of $\Delta S$ after which only a small change in $\Delta S$ is encountered. Such behavior appears to be similar to effects, obtained in real materials~\cite{2014:Ericsson:Gd-Co-Al,2015:Ericsson:Gd56Ni15Al27Zr2,2016:Ericsson:Gd50Co45Fe5,2017:Ericsson:Gd-Ni-Al,2018:Ericsson:GdMnSi}. Figure~\ref{fig:dS_U=3_lower}(c) represents results for $n = 0.66$ and as we can see in Fig.~\ref{fig:dS_U=3_lower}(c4), the rapid gap growth of $\Delta S$ is completely due to $\Delta s_{\rm{PM/AFM}}$ exhibiting strong temperature dependence.
Upon further increase in temperature a complex interplay between phase volumes and specific entropy changes results in a further increase of $\Delta S$ resulting in MCE peak.
Since in finite magnetic field the filling line for $n = 0.66$ goes right through the middle of the dome, turning on magnetic field leads to a large jump of phase volumes, Fig.~\ref{fig:dS_U=3_lower}(c3). This leads to a rapid fall of $|\Delta S|$ in PPS-subregion. It is worthwhile to note that here PPS demonstrates a rather different form compared to the previous results: at $h = 0$ homogeneous AFM phase is encountered, but an application of magnetic field causes the system to separate in ``AFM + Fi''. 
Note, that here ferrimagnetic phase does not exhibit inverse MCE and is not relevant to the overall behavior of $\Delta S$.

\section{Conclusions}\label{sec:conclusions}

The proposed research of magnetocaloric effect in metallic systems, exhibiting first order magnetic phase transition between magnetically ordered (ferro- and ferrimagnetic) and non-magnetic (antiferro- and paramagnetic) states indicates that the account for different types of magnetic phase separation (AFM + ferrimagnetic, FM + ferrimagnetic, PM + ferrimagnetic) drastically changes the behavior of the magnetocaloric effect and the temperature dependence of its characteristics. In phase-separated region the following statements hold: (i)~volume of individual phases depend on magnetic field, temperature and electron filling; (ii)~phase separation boundaries depend on magnetic field; (iii)~there exist two critical temperatures, namely the exit temperatures from phase separation in zero and finite magnetic field. From these simple statements new mechanisms of MCE emerge.
Separated phases have different response (often demonstrating opposite signs) to a magnetic field and the response becomes proportional to the phase volume. Therefore, the observable inverse MCE is proportional to a phase volume of a particular phase inside PS that exhibits positive entropy response to a magnetic field.
For example, the domination of ferrimagnetic phase that often exhibits negative response $dm/dh$ on a change in magnetic field (even though it does contradict fundamental thermodynamic equations\cite{landau2013statistical}), leads to the one-sided peak $\Delta S > 0$. Similar effect is observed in MnRhAs~\cite{ferri_anom} exhibiting ferrimagnetic order and Mn$_{0.8}$Fe$_{0.2}$Ni$_{1-x}$Cu$_x$Ge~\cite{XIAO2018916}(from FM hexagonal phase to AFM orthorhombic phase), Gd$_5$Si$_x$Ge$_{4-x}$~\cite{anom_last} (magnetostructural phase transition from PM to FM).

A particular interest is drawn towards the first exit temperature, when system transitions into a state of partial phase separation, see Fig.~\ref{fig:general}. At this temperature the entropy change exhibits a kink (sometimes with $\Delta S > 0$) and severely changes its temperature dependence: if PPS-region has PS in $h = 0$, then we encounter a linear growth of $\Delta S$ on the whole temperature interval of PPS; if PS in $h \neq 0$, then there is a rapid decrease of $\Delta S$ on the same temperature interval. Both types of PPS are important to study, since an application of magnetic field either destroys phase separation or induces it.

In the explored parameter values, we encountered that one of the phases participating in phase separation exhibits a second order magnetic transition from paramagnetic to antiferromagnetic phase.
There are two critical temperatures for the transition, namely temperature in zero and finite magnetic field. The region between transition temperatures favours a rapid growth of $\Delta S$. It represents another link to experiment, namely the tablelike behavior of the magnetocaloric effect.


Despite the fact that antiferromagnetic order was thought to be an important factor of inverse MCE, we conclude that its role is to stabilize first order magnetic transition and it can easily be interchanged with some other kind of ordering such as structural, spontaneous spin reorientation transition \cite{spin_reor_double_peak}, or order--disorder transition. Thus, any interaction may serve as a trigger for FOMT resulting in phase separation. 
On the other hand, obtained results give rigorous answer to the question of magnetic field and temperature phase volumes dependence~\cite{PS_material_anom,wrong_maxwell_best,maxwellwr_newmeth,phase_weights, Das_2010} and how it impacts MCE. 

The one-sided peak and phase separation can explain standard two-peak structure in metallic magnets, which experience first-order phase transition \cite{ferri_compound_inverse, badmax_newmeth, two_peak}. It also points out the artifacts of the use of models, which do not account for such dependencies in experimental works \cite{ferri_anom, CARON20093559, two_peak}.


Even though the chosen mean-field approximation has a few problems, such as an overestimation of critical temperature of magnetic transition, and an impossibility to account for
collective excitations and longitudinal local fluctuations, it still can qualitatively describe a system with non-homogeneous states for a problem of magnetocaloric effect. To estimate the relevant temperature scale for our results we set $D\sim 0.5-1$~eV, which is a typical value for metals. So the temperature $T = 0.1$ in our calculations is about $550-1100$ K. It is a well-known fact that the mean-field approximation tends to overestimate the critical temperature, which implies a possible proximity of the room temperature to the temperature region of the obtained phenomena.

Compound-specific results obtained in the context of MCE in the density functional approach are of high value but it can be difficult to derive any qualitative and unbiased conclusions from these investigations \cite{ferri_compound_inverse,Ni-Co-Mn-Ti,GMCE_DOS,2020:GdTX}. However, the model-based approach used in the paper, which ignores any specific features of the density of states, captures the fundamentals about metals and allows for a completely general description of the phase separation in a first order magnetic phase transition and how it impacts MCE. The approach combined with the proper density calculation of electronic structure can directly, but technically tedious, be applied to the investigations of any specific compounds for further considerations.

In the future, the obtained results can be studied using alternative approach based on another approximation applied to any itinerant electron model that does not have problems of our approximation, for example self-consistent spin fluctuation theory~\cite{1985:Moriya_book}, dynamic mean-field theory~\cite{2006:Oudovenko}. Independent of the used approximation, qualitative consequences of phase separation in the context of MCE are expected to stay the same.

Another issue is connected with an application of ``thermodynamic'' approach to the formation of phase separation in metal, which ignores actual form of phase regions within a sample and a problem of accounting for long-range Coulomb interaction. We believe that if one does not ignore this problem, then one will have a slightly reduced effect of phase separation, see account of phase region geometry in~Refs.~\onlinecite{LorenzanaI,LorenzanaII}.

\section{Acknowledgments}
P.A.I. is grateful to A. A. Katanin and V. Yu. Irkhin for
fruitful discussions. The research was carried out within the
state assignment of Ministry of Science and Higher Edu-
cation of the Russian Federation (theme “ Quantum” No.~AAAA-A18-118020190095-4), supported in part by RFBR (project
No. 20-02-00252).

\end{document}